\documentclass[amssymb, amsmath,superscriptaddress,nobibnotes,nofootinbib,floatfix,aps,prl,citeautoscript]{revtex4}
\usepackage[pdftex]{color,graphicx}

\usepackage{bm}%
\usepackage{graphicx,xspace}
\usepackage{epstopdf}

\newcommand{\ILM}{Institute of Light and Matter, UMR5306 Universit\'e Lyon 1-CNRS, Universit\'e de Lyon 69622 Villeurbanne cedex, France}
\newcommand{\ESRF}{ESRF-The European Synchrotron, CS 40220, 38043 Grenoble Cedex 9, France}
\begin{document}
\author{V. M. Giordano}
\email{valentina.giordano@univ-lyon1.fr}
\affiliation{\ILM}
\author{B. Ruta}
\email{ruta@esrf.fr}
\affiliation{\ESRF}
\date{\today}

\title{Unveiling the structural arrangements responsible for the atomic dynamics in metallic glasses during physical aging} 


\maketitle
\newpage
\textbf{Understanding and controlling physical aging, i.e. the spontaneous temporal evolution of out-of-equilibrium systems, represents one of the greatest tasks in material science. Recent studies have revealed the existence of a complex atomic motion in metallic glasses, with different aging regimes in contrast with the typical continuous aging observed in macroscopic quantities. By combining dynamical and structural synchrotron techniques, for the first time we directly connect previously identified microscopic structural mechanisms with the peculiar atomic motion, providing a broader unique view of their complexity. We show that the atomic scale is dominated by the interplay between two processes: rearrangements releasing residual stresses related to a cascade mechanism of relaxation, and medium range ordering processes, which do not affect the local density, likely due to localized relaxations of liquid-like regions. As temperature increases, a surprising additional secondary relaxation process sets in, together with a faster medium range ordering, likely precursors of crystallization.}\\

\large{\textbf{Introduction}}\\
A widespread use of out-of-equilibrium materials is at this day limited by the lack of a detailed understanding of the mechanism ruling their physical aging. This comprehension is for instance fundamental for a proper exploitation of the outstanding mechanical, physical and chemical properties of metallic glasses (MG)~\cite{zhang2006NatMat,Wang_ProgMatSci2012}. This goal requires a description of the structural and dynamical changes occurring at the atomic level which is still missing due to limitations in both experiments and numerical simulations.  
Many are the  X-ray Diffraction (XRD) studies which have investigated the subtle structural changes occurring during aging induced by annealings close to the glass transition temperature $T_{\rm g}$~\cite{Yavari_ActaMat2005,Louzguine_JAllComp2007,Bednarcik_JAllComp2010,Bruning_PRB1990,Bednarcik_JPCM2011, Mao_PRB1995,Evenson_PhysRevB2014, Scudino_JAllComp2015}. The existence of two orthogonal microscopic mechanisms has been highlighted: an irreversible atomic rearrangement, associated with density changes and radial atomic motion observed only in as-cast glasses, quenched with a cooling rate as high as $10^6$~K s$^{-1}$ (hyper-quenching), and a reversible one, not affecting the density, but changing the short range order, taking place also in glasses annealed above $T_{\rm g}$.
The former process is usually ascribed to "free volume" or "structural defects" annihilation \cite{taub1980,Li2013}. Such structural defects have been described as centers of internal strain due to the trapping of excess empty space during quenching~\cite{Egami_JMatSci1978} or related to the presence of icosahedral clusters, causing geometrical frustration in the glass~\cite{Shen_PRL2009} or again to density inhomogeneities on the nano-scale~\cite{Scudino_JAllComp2015}. 

The recent introduction of X-ray Photon Correlation Spectroscopy (XPCS) for following the atomic motion in hard materials~\cite{Ruta_NatComm2014,Ruta_PRL2012,leitner2009} 
has directly proved the existence of such atomic rearrangements in MG even for temperatures far below $T_{\rm g}$. The observed dynamics is characterized by an unexpected local ergodicity in the glass with a "compressed" decay of the density-density correlation function, $f(q,t)$, which strongly disagrees with the typical stretched exponential shape observed in supercooled liquids~\cite{Berthier_RevModPhys2011}. A similar 
behavior has been recently associated to elastically interacting activated events in yield stress and jammed materials~\cite{Ferrero_PRL2014, Cipelletti_FarDisc2003,bouchaud2001}
 and could be related to the existence of microscopic elastic heterogeneities~\cite{Dmowski2010,Wagner_NatMat2011} and atomistic free volume zones ~\cite{ye2010NatMat}. This behavior suggests the existence of a complex mechanism ruling the atomic dynamics, previously unreported in macroscopic studies~\cite{Wang_ProgMatSci2012} or in any current theory for glasses~\cite{Berthier_RevModPhys2011,lubchenko2004}. In addition, all studies performed so far on hyper-quenched MG show that the structural relaxation process displays distinct dynamical regimes at the atomic level and then becomes stationary \cite{Ruta_PRL2012, Ruta_JCP2013,ruta2013b}, even if macroscopically the system is still aging in a continuous way~\cite{Wang_ProgMatSci2012,pineda2013}~(see Supplementary Figure 1 in Supplementary Note 1 for additional explanation).\\

In order to clarify the unique atomic motion of MG, here we combine time-resolved XPCS and high energy XRD for a thorough investigation of the microscopic structural and dynamical changes occurring during annealing in a rapidly quenched ${\rm Pd_{77}Si_{16.5}Cu_{6.5}}$ MG ($T_{\rm g}=625 K$~\cite{pineda2010}). Our results allow us to uniquely ascribe the distinct features of the atomic motion to an interplay between defects or stresses annihilation and increasing medium range order not involving density changes. 
We observe also the surprising thermal activation of a secondary relaxation process at high temperature in the decay of the density fluctuations, not reported in previous studies, which appears related to the onset of the crystallization.\\ 

\large{\textbf{Results}}\\
\textbf{Atomic motion.} XPCS data were collected at the position of the first sharp diffraction peak (FSDP), $q_0=2.81$ \AA$^{-1}$, during long subsequent isotherms at selected temperatures below $T_{\rm g}$. Figure \ref{Fig1}~(a) shows intensity-intensity correlation functions measured during annealing at $T=453 K$. The decay provides information on the dynamics, being $g_2(t)$ related to the density fluctuations through $g_2(t)-1=c\left|f(q,t)\right|^2$ where $c$ is a set-up dependent contrast~\cite{ruta2013b} and $f(q,t)$ the intermediate scattering function.
Lines are fits with the Kohlrausch-Williams-Watts (KWW) model $g_2(t)-1=c\cdot f_q^2(t_{\rm a},T) \exp[-2(t/\tau_{\alpha}(t_{\rm a},T))^{\beta(t_{\rm a},T)}]$, where all parameters depend both on temperature and on the annealing time,$t_a$, at any given $T$:
$\tau_{\alpha}(t_{\rm a},T)$ is the structural relaxation time, $\beta(t_{\rm a},T)$ the shape parameter and $f_q(t_{\rm a},T)$ is the nonergodic plateau of $f(q,t)$ associated to the trapping of the particles in the nearest neighbors cage before their escape during the $\alpha$ relaxation~\cite{cavagna2009}. 
\\
The full decorrelation to zero of $g_2-1$ confirms the existence of atomic rearrangements on a scale $2\pi/q_0=2.2$~\AA~. 
As in Refs.~\cite{Ruta_PRL2012, Ruta_JCP2013} this local ergodicity does not mean that the system relax into an equilibrium state but simply that it can evolve through different microscopic glassy configurations.
During the annealing the system ages as signaled by the continuous shift of the decay time toward larger values on increasing $t_{\rm a}$. 
This is the fast aging regime observed also in other hyper-quenched MG~\cite{Ruta_PRL2012, Ruta_JCP2013,ruta2013b}: for all $T< 513 K$ $\tau_{\alpha}$ increases exponentially during the isotherm while abruptly decreases during $T$ changes, due to the increasing thermal motion (Fig.~\ref{Fig1}~(b)). For each $T$, its behavior can be described by the empirical law $\tau_{\alpha}(t_{\rm a},T)=\tau_0(T)exp(t_{\rm a}/\tau^{\ast})$, with $\tau^{\ast} \sim 6000 $ s almost $T$ independent and close to previous reports~\cite{Ruta_PRL2012, Ruta_JCP2013}. 
At $T=513 K$ the aging abruptly stops even if the system is still far from $T_{\rm g}$ ($T/T_{\rm g}=0.79$): $\tau_{\alpha}(t_{\rm a},T)$ remains constant, and the glass enters a second stationary regime where the correlation functions still decay without however exhibiting any dependence on $t_{\rm a}$ on the probed experimental timescale. 
\\
Both dynamical regimes firmly contrast with the steady aging of macroscopic quantities in glasses~\cite{Wang_ProgMatSci2012,pineda2013}
.  While the fast aging has been previously reported for jammed soft materials~\cite{Cipelletti_FarDisc2003}, the second one is completely at odds with previous works~\cite{Berthier_RevModPhys2011}. 
Interestingly, the crossover between the two aging regimes is here accompanied by a sudden reduction of the shape parameter $\beta(t_{\rm a},T)$ from $1.82\pm0.09$ for $T<513K$ to $1.31\pm0.08$ at $T=513K$ (Fig.~\ref{Fig1}~(c)). In the optics of a stress dominated dynamics, as the one characterized by compressed ($\beta>1$) correlation functions~\cite{Ruta_PRL2012}, this decrease can be viewed as an abrupt release of internal stresses or defect annihilation~\cite{Li2013}stored in the system at low $T$ and it's indeed in agreement with the occurrence of a dynamical transition. 

Even more interesting is the temperature dependence of the initial nonergodic plateau of $g_2(t)-1$, $c \cdot f_q^2(t_{\rm a},T)$. As shown in Fig.~\ref{Fig1}~(d), all data measured for $T < 493 K$ collapse on a single curve, when reported as a function of $t/ \tau_{\alpha}(t_{\rm a},T)$. The validity of this superposition principle confirms the existence of a unique dynamical regime at low $T$. The scaling breaks at $T=493K$ due to a dramatic drop of $c \cdot f_q^2(t_{\rm a},T)$ of $\sim17\%$ in only 20 K (Fig.~\ref{Fig1}~(e)). In the macroscopic limit of $q\rightarrow 0$, the long-time plateau $f_q(t_{\rm a},T)$ is related to the elastic properties of the material and it usually steadily decreases with increasing temperature due to a reducing stiffness~\cite{ruta2012b}. This argument is however not valid in hyper-quenched MG where the first annealing leads usually to an increase in the elastic modulus~\cite{Hachenberg_ApplPhysLett2008}. In addition the 
$T$ dependence of the nonergodicity level should disappear at the inter-particle distances probed in this work~\cite{ruta2012b}. Curiously, the observed decrease occurs before the dynamical crossover in $\tau_{\alpha}(t_{\rm a},T)$ and $\beta(t_{\rm a},T)$ (Fig.~\ref{Fig1}~(b) and (c)) and persists with temperature. This early signature of the dynamical transition can be understood as the surprising thermal activation of a second decay step in the correlation function, before the $\alpha$-process, with a relaxation time, $\tau_{\beta}$, too small to be directly observed in our $g_2(t)-1$, but with a strength which leads to a $\sim33\%$ total reduction of the nonergodic level associated to the main $\alpha$-relaxation (Fig. 2).

A similar low initial plateau has been reported for few 
gels and colloidal suspensions \cite{orsi2014,guo2011,czakkel2011} 
but, to the best of our knowledge, never for structural glasses. Indeed there exist just weak 
signatures of secondary $\beta$ relaxation processes in the $f(q,t)$ in glass-formers~\cite{sidebottom1993,mezei1987}. These processes are usually observed through mechanical and dielectric spectroscopy measurements~\cite{ngaibook}. The most renowned is the Johari-Goldstein (JG) $\beta$ relaxation, reported in all types of glass formers~\cite{johari1970,Yu_MatTod2013,Liu_NatComm2014}, which behaves in a symbiotic way with the main $\alpha$-process~\cite{ngaibook}. Its occurrence has been however never observed in the temporal decay of $f(q,t)$, likely due to its very weak signature in the time domain accordingly to its broad shape in the frequency range. The JG relaxation has been reported in mechanical studies of our MG~\cite{Hachenberg_ApplPhysLett2008}, however we rule out a possible connection with the process here observed, because at the lowest heating rate, 0.5~K min$^{-1}$, its activation temperature is lower than the one here found and would still decrease if long isotherms were performed. If present, the JG should be already active at the lowest investigated $T$ and thus cannot be identified with the observed thermally activated process. The existence of fast additional $\beta$-processes beyond the well-known JG ones has been recently reported also through mechanical measurements in La-based MG~\cite{wang2014}, and ascribed to the presence of a variety of mobile atomic pairs active at low temperatures which then merge with the $\alpha$-process at high $T$.
Our process is thus still different, as it appears at high temperatures, together with the $\alpha$ one. \\

\textbf{Structural evolution.} In order to get insight on the atomic rearrangements behind the observed dynamics, we have investigated by XRD the evolution of the $S(q)$ following the same thermal protocol. Previous works have shown that an accurate investigation of position and shape of the FSDP allows to track the subtle structural changes associated to structural defects annihilation in hyper-quenched MG~\cite{Yavari_ActaMat2005, Ruta_JAllComp2014,Evenson_PhysRevB2014}. However, the overall view can be obtained only through a real-space analysis, for which large-angle high quality data are necessary~\cite{Scudino_JAllComp2015}. We have thus performed two experiments, with a setup allowing to investigate a $q$ range around the FSDP with high resolution, and, successively, a setup allowing to access up to $q \approx 25$ \AA$^{-1}$ in order to extract the pair distribution function $G(r)$. 
\\
Fig.~\ref{Fig2} reports data collected at the beginning of each isotherm. 
The most direct information comes from the fitting of the FSDP, that we model with an asymmetric pseudo-Voigt function~\cite{Ruta_JAllComp2014}. 
The temperature evolution of the position $q_0$ of the FSDP has been largely used for tracking volume changes~\cite{Yavari_ActaMat2005,Louzguine_JAllComp2007,Bednarcik_JAllComp2010}: 
$(q_0(298K)/q_0(T,t_{\rm a}))^3=V(T,t_{\rm a})/V_0$, with $V_0$ the room temperature volume. The validity of this approach has been contested due to the different temperature behavior of the other maxima of $S(q)$~\cite{Mattern_ActaMat2012}, but it has been confirmed very recently and explained with the direct connection of FSDP to the medium range order (MRO)~\cite{Scudino_JAllComp2015}. 
In our case, the volume thermal expansion coefficient obtained by the relative $q_0$ change, $\alpha_{\rm q}=(3.5 \pm 0.1) \times 10^{-5} K^{-1}$, is in very good agreement with the macroscopic coefficient~\cite{Qian_MatSciEng1991}, confirming the validity of this approach for our MG.
Fig.~\ref{Fig3}~(a) reports the temperature and time evolution of $V/V_0$: for each isotherm a slight volume reduction is observed with time, which can be interpreted as a structural defects (or density inhomogeneities) annihilation, and can be fitted with the exponential law  $x(t_{\rm a}) = (x_0-x_{\infty}) e^{-t_{\rm a}/\tau_{\rm V}}+x_{\infty}$, where $x(t_{\rm a})=V(T,t_{\rm a})/V_0$ and $\tau_{\rm V}$ is the characteristic time at each $T$.  At constant temperature, the only phenomenon taking place is aging: 
no side effects, such as thermal expansion or temperature-induced change in the atomic mobility, are present. As temperature increases the total isothermal densification decreases, as shown in Fig.~\ref{Fig3}~(b). 
At the highest investigated $T$, 513~K, there is no further volume reduction, corresponding to the full structural defects annihilation. 
Such complete release is concomitant with the crossover between aging regimes observed with XPCS (Fig.~\ref{Fig1}~(b)), suggesting that structural defects annihilation is responsible for the fast aging. However, while they are of the same order of magnitude, the aging rate $\tau^{\ast}$ and $\tau_{\rm V}$ display a different $T$ dependence, being the first one almost $T$ independent whereas the latter decreases with $T$, accordingly with a higher atomic mobility (Fig.~\ref{Fig3}~(d)). \\
As shown in Fig.~\ref{Fig3}~(c) the densification is accompanied by a continuous narrowing of the FSDP, which exhibits aging itself and goes on even when there is no further densification. It has been shown that the FSDP in MG is mostly related to the atomic arrangements for $r \ge 6$ \AA~\cite{Ma_NatMat2009}, and its width has been often used to calculate the correlation length over which the period of a repeated unit survives~\cite{Sokolov_PRL1992}. Its narrowing is thus an indicator of an increasing medium range order.
A fit of the relative width $\Gamma/\Gamma_0$ with the same exponential law allows us to obtain the characteristic time for this narrowing, $\tau_{{\rm \Gamma}}$, which, inversely from $\tau_{\rm V}$, increases with $T$ (Fig.~\ref{Fig3}~(d)). This slowing down can be understood in terms of a reduced ability for the atoms to rearrange because of the concomitant densification. At the same time, the total isothermal change in $\Gamma/\Gamma_0$ increases with $T$, indicating that the ordering becomes more efficient. 
If we take the average of the two characteristic times, $\tau_{\rm V}$ and $\tau_{{\rm \Gamma}}$, the agreement with $\tau^{\ast}$ is impressive (Fig.~\ref{Fig3}~(d)).
We can thus associate the fast dynamical aging to both processes: a structural defects annihilation and a medium range ordering, as far as they affect density. 
Once the structural defects are completely annihilated, no further density change can take place, thus XPCS does not see aging and we enter in the stationary regime.
Here, $\tau_{\alpha}(T,t_{\rm a})$ is still many orders of magnitude  larger than the extrapolated value in the supercooled liquid~\cite{pineda2010}, confirming that the system is still a glass, and the structural relaxation takes place through an ordering mechanism which does not affect density anymore. \\
As reported above, the crossover between the two dynamical regimes is accompanied by a sudden decrease of the shape parameter $\beta(T,t_{\rm a})$, which however, remains larger than 1. This means that while residual stresses due to the presence of structural defects have been annihilated, other stresses still exist with a different origin, likely related to a frustration in the repetition and ordering of the atomic clusters, similarly to the structural rigidity suggested for network glasses by topological constraints models~\cite{greaves2007}. This result suggests a discontinuous release of stresses in MG at the atomic level although a steady volume reduction.  \\
Finally, starting from T=493~K, the FSDP narrowing becomes sharper, revealing a more important ordering process. This is also the activation temperature of the secondary faster relaxation process observed with XPCS. 
In order to get a deeper understanding of this latter, we have analyzed the position and width of the neighbors shells in G(r) up to $13 $\AA:  Fig.~\ref{Fig4} reports them for the peaks located at $r_{\rm 1}=2.79$ \AA, $r_{\rm 2}=4.67$ \AA  ~(the peak at $r_{\rm 3}=5.34$ \AA~  changing accordingly), and $r_{\rm 4} = 7 $ \AA.
Tiny changes due to aging are beyond our resolution, except during the first isotherm. Here we observe the shrinking of the nearest neighbors shell and expansion of the next-nearest neighbors one, corresponding to the simultaneous annihilation of n-type (denser) and p-type (less dense) defects, distributed on different length scales, as recently reported~\cite{Scudino_JAllComp2015}. Successive heating leads to the thermal expansion of the first two shells, and the increase of the nearest neighbors width due to thermal vibrations. \\
The most interesting behavior is the one of the third shell ($r_{\rm 4}$), corresponding to the medium range and more directly connected to the FSDP. Here we observe a continuous narrowing, while the position is almost $T$ independent up to 493~K, where it suddenly starts to increase. 
We can thus understand the secondary relaxation as due to the thermal activation of atomic rearrangements in the medium range, favoring an increase in the ordering correlation length for $T/T_{\rm g} \ge 0.79$: this suggests that it is a precursor of crystallization which indeed starts at slightly higher temperatures (see Supplementary Note 2). \\ 
\\
\large{\textbf{Discussion}}\\
Our study highlights the importance of performing a simultaneous accurate investigation of both dynamical and structural features in order to obtain the overall view of the complex scenario occurring at the atomic level in metallic glasses. We find that the subtle structural changes usually observed by XRD studies~\cite{Yavari_ActaMat2005,Louzguine_JAllComp2007,Bednarcik_JAllComp2010,Bruning_PRB1990,Bednarcik_JPCM2011, Mao_PRB1995,Evenson_PhysRevB2014, Scudino_JAllComp2015} give raise to dramatic dynamical features as revealed by XPCS, like a peculiar fast aging at low temperatures or the thermal activation of a surprising secondary relaxation process at high temperature. In particular our work shows that we do observe aging in the atomic motion only if there is a simultaneous structural change affecting the density. This correlation between dynamics and structure is likely independent on the precise details of the aging which instead could depend on the history or composition of the glass. \\
Figure \ref{Fig6} shows a schematic view of our results. The decay of the intensity correlation functions obtained with XPCS (panel a) can be associated to particles movements in the probed sample volume. In a very approximate way, the $g_2(t)$ can therefore be viewed as a measurement of the overlap between an initial state of the glass (blue circles in the squared box) and its temporal evolution (green circles in the squared box). Looking now to the two aging regimes observed in this work, we can interpret the fast aging, associated to density changes, as the one taking place when particles move so much that the number of particles in the probed volume actually changes (density change, squares in panel (b)). During this regime the decay of the $g_2(t)$ shifts also toward longer time scales with annealing time and consequently $\tau_{\alpha}$ increases (see Fig. \ref{Fig1}~(a)).\\
Differently, when density inhomogeneities are fully annihilated, the number of particles in the probed volume is constant, but still they slightly move, thus the glass changes configuration and becomes more ordered while keeping the same density (medium range ordering, squares in panel (c)). From the dynamical point of view this regime corresponds to a constant structural relaxation time $\tau_{\alpha}$ (Fig. \ref{Fig1}~(b)).\\
The typical aging observed at macroscopic scales \cite{Wang_ProgMatSci2012} suggests that the system still stiffens with time independently on the actual microscopic mechanism (Fig. \ref{Fig6}~(b) and (c)). 
\\
From the point of view of the potential energy landscape (PEL)~\cite{debenedetti2001,rodney2011,heuer2008}, the fast aging regime corresponds to the thermal activation of a cascade of jumps from a high energy local minimum to a deeper relaxed state, in agreement with the avalanches-based relaxation recently reported in numerical simulations and mechanical measurements~\cite{fan2014,fan2015,samwer2014}.
Once all the stresses related to the hyper-quenching and corresponding to density defects are released, the material is in a sort of “relaxed denser state  relatively to the experimental thermal protocol”. This does not mean that all stresses are annihilated as their presence is still confirmed by the compressed decay of the correlation curves.
At this stage, the system is trapped in a local minimum of the PEL in a macroscopically relaxing matrix in agreement with the localized dynamics reported in Ref.~\cite{fan2015}, suggesting the existence of liquid-like regions~\cite{wang2014Natcom}. We are in the regime where no dynamical aging is observed anymore at the atomic scale even if macroscopic observables still evolve continuously with time toward the equilibrium liquid values~\cite{Wang_ProgMatSci2012,pineda2013,busch1998b}.\\
The evolution from the first to the second relaxation mechanism could be read as a ductile to brittle transition~\cite{Kumar2013,Kumar2009,Kumar2011b}. Indeed the as-cast MG, with continuous irreversible atomic rearrangements associated to density changes, appears as very ductile, with extremely large embrittlement times. The disappearance of the first mechanism suggests a 
loss of ductility and the possibility of annealing-induced embrittlement in agreement also with the brittle behaviour of Pd-based MG for low cooling rates~\cite{Kumar2011b} . 
To confirm this interpretation, however, further mechanical investigations are required, which are out of our scope.\\

In our study we report also the first direct experimental observation of an additional relaxation process in the temporal evolution of the $f(q,t)$, with a decay time much faster than that of the main structural relaxation. 
It is important to underline that a secondary process does not involve the motion of the entire matrix, which remains governed by the last decay  (structural relaxation), still slow and with a characteristic time of the order of a thousands of seconds at the atomic level. 
We relate this secondary process to an incipient crystallization as its thermal activation coincides with a marked increase in the medium range ordering and a dilatation of the third shell. Its activation means that on increasing temperature, some fast degrees of freedom frozen in the deep glassy state are reactivated, likely related to the onset of nucleation or to a phase separation prior to crystallization, as it occurs just few temperature steps below the detection of incipient crystallization in the XRD spectra and the corresponding increase in the static component of the XPCS data (see Supplementary Note 2). Of course the confirmation of this interpretation requires further investigations which are beyond the purpose of this work. \\ 
It is worth underlying that these results provide a direct connection between dynamical and structural microscopic evolutions in MG
which is fundamental for developing a microscopic theory for aging and ultimately design new amorphous materials with improved stability. 
Finally, the understanding of the atomic motion and aging in MG opens also the way to the comprehension of similar mechanisms in complex systems like jammed soft materials and many biological systems, being glasses often considered as archetypes of out-of-equilibrium systems~\cite{Liu_Nature1998}.\\
 
\large{\textbf{Methods}}\\
{\textbf{Sample preparation.}}
The samples were produced as thin ribbons by arc melt spinning the pure elements at the Polytechnic University of Catalonia as explained in Ref. \cite{pineda2010}. The resulting metallic ribbons have a thickness of $15\pm4 \mu m$, close to the optimal value for maximizing the scattered intensity while keeping a reasonable contrast in the XPCS experiments. \\

{\textbf{X-ray photon-correlation spectroscopy.}}
XPCS experiments were performed at the ID10 beamline at ESRF in Grenoble, France, by using an incident X-ray beam wavelength $\lambda=1.55$ \AA, and a coherent flux of $\sim 10^{10}$ photons per second per $200$ mA. The experimental setup and the data treatment are discussed in detail in Refs.~\cite{Ruta_NatComm2014,Chushkin2012}. 
The dynamics was measured during long isotherm steps at different selected temperatures in the glass while $T$ was increased with a fixed rate of 3K min$^{-1}$. At every $T$, the $S(q)$, was measured as well and the reproducibility of the results has been checked in a second experiment.\\

{\textbf{X-ray diffraction measurements.}}
XRD experiments were performed at the ID15 beamline at ESRF with an incident wavelength $\lambda=0.1425 $\AA$^{-1}$. Up to 5 ribbons were put together for improving the scattered intensity and the signal to noise ratio.
The first experiment was done using a 2D Pixium detector located far from the sample at an angle corresponding to the position $q_0$ of the FSDP, in order to maximize the resolution. In the second experiment a 2D MAR CCD detector was used instead, centered with respect to the incident beam and located close to the sample in order to access up to $q\approx 25$\AA and to be able to calculate with a good resolution the pair distribution function $G(r)$.\\

{\textbf{X-ray diffraction data analysis.}}
Experimental intensities have been normalized and reported to absolute units following a standard procedure~\cite{Ruta_JAllComp2014}. 
More in detail, assuming independent atomic contributions to the scattering, $S(q)$ is calculated as:
\begin{equation}
S(q)=\frac{\alpha I_{\rm s} (q)-I_{\rm C} (q)}{Z_{{\rm tot}}^2 f_{{\rm eff}}^2 (q)}
\end{equation}
Where $Z_{{\rm tot}}$ is the sum, weighted by the atomic concentrations, of the atomic numbers $Z$ of the atoms, $I_C$ is the Compton scattering calculated as the sum of the individual atomic components and $f_{{\rm eff}}$ is the effective electronic form factor, calculated as the weighted sum of the atomic scattering factors, divided by $Z_{{\rm tot}}$. $I_{\rm s}$ is the sample scattering, after subtraction of the empty cell and $\alpha$ is a density-dependent normalization factor, which is found following the iterative procedure reported in Ref.~\cite{Eggert2002}.\\
The 2D images have been integrated in slices parallel and perpendicular to the ribbon axis in order to check the presence of structural anisotropy and its possible evolution with time and temperature. The results of this analysis, shown in the Supplementary Note 3, reveal that a slight anisotropy does exist but does not evolve during the whole thermal protocol.
The $G(r)$ has then been calculated by a sine Fourier transformation of $S(q)$ for $q \le 13$ \AA$^{-1}$, the higher $qs$ being too noisy:
\begin{equation}
G(r) = 4 \pi r \rho [g(r) - 1] = \frac{2}{\pi} \int{q \left[S(q)-1\right] sin\left(qr\right) dq}
\label{eqGr}
\end{equation}
where $g(r)$ is the atomic pair distribution function and $\rho$ the average atomic density. 

The first shell, located at $r_{\rm 1}=2.79$ \AA, has been fitted using an asymmetric gaussian function, while all the other shells using from 2 to 4 gaussian functions.


\begin{thebibliography}{10}
\expandafter\ifx\csname url\endcsname\relax
  \def\url#1{\texttt{#1}}\fi
\expandafter\ifx\csname urlprefix\endcsname\relax\def\urlprefix{URL }\fi
\providecommand{\bibinfo}[2]{#2}
\providecommand{\eprint}[2][]{\url{#2}}

\bibitem{zhang2006NatMat}
\bibinfo{author}{Zhang, Y.}, \bibinfo{author}{Wang, W.~H.} \&
  \bibinfo{author}{Greer, A.~L.}
\newblock \bibinfo{title}{Making metallic glasses plastic by control of
  residual stress}.
\newblock \emph{\bibinfo{journal}{Nat. Mat.}} \textbf{\bibinfo{volume}{5}},
  \bibinfo{pages}{857} (\bibinfo{year}{2006}).

\bibitem{Wang_ProgMatSci2012}
\bibinfo{author}{Wang, W.~H.}
\newblock \bibinfo{title}{The elastic properties, elastic models and elastic
  perspectives of metallic glasses}.
\newblock \emph{\bibinfo{journal}{Prog. Mater. Sci.}}
  \textbf{\bibinfo{volume}{57}}, \bibinfo{pages}{487--656}
  (\bibinfo{year}{2012}).

\bibitem{Yavari_ActaMat2005}
\bibinfo{author}{Yavari, A.~R.} \emph{et~al.}
\newblock \bibinfo{title}{Excess free volume in metallic glasses measured by
  x-ray diffraction}.
\newblock \emph{\bibinfo{journal}{Acta Mat.}} \textbf{\bibinfo{volume}{53}},
  \bibinfo{pages}{1611--1619} (\bibinfo{year}{2005}).

\bibitem{Louzguine_JAllComp2007}
\bibinfo{author}{Louzguine-Luzgin, D.} \emph{et~al.}
\newblock \bibinfo{title}{Free volume and elastic properties changes in
  cu׺r״iװd bulk glassy alloy on heating}.
\newblock \emph{\bibinfo{journal}{J. Alloys Comp.}}
  \textbf{\bibinfo{volume}{431}}, \bibinfo{pages}{136} (\bibinfo{year}{2007}).

\bibitem{Bednarcik_JAllComp2010}
\bibinfo{author}{Bednarcik, J.}, \bibinfo{author}{Curfs, C.},
  \bibinfo{author}{Sikorski, M.}, \bibinfo{author}{Franz, H.} \&
  \bibinfo{author}{Jiang, J.}
\newblock \bibinfo{title}{Thermal expansion of La-based BMG studied by in situ
  high-energy x-ray diffraction}.
\newblock \emph{\bibinfo{journal}{J. Alloys Comp.}}
  \textbf{\bibinfo{volume}{504S}}, \bibinfo{pages}{155} (\bibinfo{year}{2010}).

\bibitem{Bruning_PRB1990}
\bibinfo{author}{Bruning, R.} \& \bibinfo{author}{Strom-Olsen, J.~O.}
\newblock \bibinfo{title}{Atomic displacements during structural relaxation in
  a metallic glass}.
\newblock \emph{\bibinfo{journal}{Phys. Rev. B}} \textbf{\bibinfo{volume}{41}},
  \bibinfo{pages}{2678} (\bibinfo{year}{1990}).

\bibitem{Bednarcik_JPCM2011}
\bibinfo{author}{Bednarcik, J.} \emph{et~al.}
\newblock \bibinfo{title}{Thermal expansion of a la-based bulk metallic glass:
  insight from in situ high-energy x-ray diffraction}.
\newblock \emph{\bibinfo{journal}{J. Phys.: Condens. Matter}}
  \textbf{\bibinfo{volume}{23}}, \bibinfo{pages}{254204}
  (\bibinfo{year}{2011}).

\bibitem{Mao_PRB1995}
\bibinfo{author}{Mao, M.}, \bibinfo{author}{Altounian, Z.} \&
  \bibinfo{author}{Bruning, R.}
\newblock \bibinfo{title}{X-ray diffraction study of structural relaxation in
  metallic glasses}.
\newblock \emph{\bibinfo{journal}{Phys. Rev. B}} \textbf{\bibinfo{volume}{51}},
  \bibinfo{pages}{2798} (\bibinfo{year}{1995}).

\bibitem{Evenson_PhysRevB2014}
\bibinfo{author}{Evenson, Z.} \emph{et~al.}
\newblock \bibinfo{title}{$\beta$ relaxation and low-temperature aging in a
  au-based bulk metallic glass: From elastic properties to atomic-scale
  structure}.
\newblock \emph{\bibinfo{journal}{Phys. Rev. B}} \textbf{\bibinfo{volume}{89}},
  \bibinfo{pages}{174204} (\bibinfo{year}{2014}).

\bibitem{Scudino_JAllComp2015}
\bibinfo{author}{Scudino, S.} \emph{et~al.}
\newblock \bibinfo{title}{Length scale-dependent structural relaxation in
  zr57.5ti7.5nb5cu12.5ni10al7.5 metallic glass}.
\newblock \emph{\bibinfo{journal}{J. Alloy and Comp.}}
  \textbf{\bibinfo{volume}{639}}, \bibinfo{pages}{465--469}
  (\bibinfo{year}{2015}).

\bibitem{taub1980}
\bibinfo{author}{Taub, A.~I.} \& \bibinfo{author}{Spaepen, F.}
\newblock \bibinfo{title}{The kinetics of structural relaxation of a metallic
  glass}.
\newblock \emph{\bibinfo{journal}{Acta Metall.}} \textbf{\bibinfo{volume}{28}},
  \bibinfo{pages}{1781--1788} (\bibinfo{year}{1980}).

\bibitem{Li2013}
\bibinfo{author}{Li, W.}, \bibinfo{author}{Bei, H.}, \bibinfo{author}{Tong,
  Y.}, \bibinfo{author}{Dmowski, W.} \& \bibinfo{author}{Gao, Y.~F.}
\newblock \bibinfo{title}{Structural heterogeneity induced plasticity in bulk
  metallic glasses: From well-relaxed fragile glass to metal-like behavior}.
\newblock \emph{\bibinfo{journal}{Appl. Phys. Lett.}}
  \textbf{\bibinfo{volume}{103}}, \bibinfo{pages}{171910}
  (\bibinfo{year}{2013}).

\bibitem{Egami_JMatSci1978}
\bibinfo{author}{Egami, T.}
\newblock \bibinfo{title}{Structural relaxation in amorphous fe40ni40p14b6
  studied by energy dispersive x-ray diffraction}.
\newblock \emph{\bibinfo{journal}{J. Mater. Sci.}}
  \textbf{\bibinfo{volume}{13}}, \bibinfo{pages}{2587--2599}
  (\bibinfo{year}{1978}).

\bibitem{Shen_PRL2009}
\bibinfo{author}{Shen, Y.~T.}, \bibinfo{author}{Kim, T.~H.},
  \bibinfo{author}{Gangopadhyay, A.~K.} \& \bibinfo{author}{Kelton, K.~F.}
\newblock \bibinfo{title}{Icosahedral order, frustration, and the glass
  transition: Evidence from time-dependent nucleation and supercooled liquid
  structure studies}.
\newblock \emph{\bibinfo{journal}{Phys. Rev. Lett.}}
  \textbf{\bibinfo{volume}{102}}, \bibinfo{pages}{057801}
  (\bibinfo{year}{2009}).

\bibitem{Ruta_NatComm2014}
\bibinfo{author}{Ruta, B.} \emph{et~al.}
\newblock \bibinfo{title}{Revealing the fast atomic motion of network glasses}.
\newblock \emph{\bibinfo{journal}{Nat. Commun.}} \textbf{\bibinfo{volume}{5}},
  \bibinfo{pages}{3939} (\bibinfo{year}{2014}).

\bibitem{Ruta_PRL2012}
\bibinfo{author}{Ruta, B.} \emph{et~al.}
\newblock \bibinfo{title}{Atomic-scale relaxation dynamics and aging in a
  metallic glass probed by x-ray photon correlation spectroscopy}.
\newblock \emph{\bibinfo{journal}{Phys. Rev. Lett.}}
  \textbf{\bibinfo{volume}{109}}, \bibinfo{pages}{165701}
  (\bibinfo{year}{2012}).

\bibitem{leitner2009}
\bibinfo{author}{Leitner, M.}, \bibinfo{author}{Sepiol, B.},
  \bibinfo{author}{Stadler, L.~M.}, \bibinfo{author}{Pfau, B.} \&
  \bibinfo{author}{Vogl, G.}
\newblock \bibinfo{title}{Atomic diffusion studied with coherent x-rays}.
\newblock \emph{\bibinfo{journal}{Nat. Mat.}} \textbf{\bibinfo{volume}{8}},
  \bibinfo{pages}{717--720} (\bibinfo{year}{2009}).

\bibitem{Berthier_RevModPhys2011}
\bibinfo{author}{Berthier, L.} \& \bibinfo{author}{Biroli, G.}
\newblock \bibinfo{title}{Theoretical perspective on the glass transition and
  amorphous materials}.
\newblock \emph{\bibinfo{journal}{Rev. Mod. Phys.}}
  \textbf{\bibinfo{volume}{83}}, \bibinfo{pages}{587--645}
  (\bibinfo{year}{2011}).

\bibitem{Ferrero_PRL2014}
\bibinfo{author}{Ferrero, E.~E.}, \bibinfo{author}{Martens, K.} \&
  \bibinfo{author}{Barrat, J.-L.}
\newblock \bibinfo{title}{Relaxation in yield stress systems through
  elastically interacting activated events}.
\newblock \emph{\bibinfo{journal}{Phys. Rev. Lett.}}
  \textbf{\bibinfo{volume}{113}}, \bibinfo{pages}{248301}
  (\bibinfo{year}{2014}).

\bibitem{Cipelletti_FarDisc2003}
\bibinfo{author}{Cipelletti, L.} \emph{et~al.}
\newblock \bibinfo{title}{Universal non-diffusive slow dynamics in aging soft
  matter}.
\newblock \emph{\bibinfo{journal}{Faraday Discuss. Chem. Soc.}}
  \textbf{\bibinfo{volume}{123}}, \bibinfo{pages}{237} (\bibinfo{year}{2003}).

\bibitem{bouchaud2001}
\bibinfo{author}{Bouchaud, J.~P.} \& \bibinfo{author}{Pitard, E.}
\newblock \bibinfo{title}{Anomalous dynamical light scattering in soft glassy
  gels}.
\newblock \emph{\bibinfo{journal}{Eur. Phys. J. E}}
  \textbf{\bibinfo{volume}{6}}, \bibinfo{pages}{231--236}
  (\bibinfo{year}{2001}).

\bibitem{Dmowski2010}
\bibinfo{author}{Dmowski, W.}, \bibinfo{author}{Iwashita, T.},
  \bibinfo{author}{Chuang, C.-P.}, \bibinfo{author}{Almer, J.} \&
  \bibinfo{author}{Egami, T.}
\newblock \bibinfo{title}{Elastic heterogeneity in metallic glasses}.
\newblock \emph{\bibinfo{journal}{Phys. Rev. Lett.}}
  \textbf{\bibinfo{volume}{105}}, \bibinfo{pages}{205502}
  (\bibinfo{year}{2010}).

\bibitem{Wagner_NatMat2011}
\bibinfo{author}{Wagner, H.} \emph{et~al.}
\newblock \bibinfo{title}{Local elastic properties of a metallic glass}.
\newblock \emph{\bibinfo{journal}{Nature Mater.}}
  \textbf{\bibinfo{volume}{10}}, \bibinfo{pages}{439} (\bibinfo{year}{2011}).

\bibitem{ye2010NatMat}
\bibinfo{author}{Ye, J.~C.}, \bibinfo{author}{Lu, J.}, \bibinfo{author}{Liu,
  C.~T.}, \bibinfo{author}{Q.Wang} \& \bibinfo{author}{Yang, Y.}
\newblock \bibinfo{title}{Atomistic free-volume zones and inelastic deformation
  of metallic glasses}.
\newblock \emph{\bibinfo{journal}{Nature Mater.}} \textbf{\bibinfo{volume}{9}},
  \bibinfo{pages}{619} (\bibinfo{year}{2010}).

\bibitem{lubchenko2004}
\bibinfo{author}{Lubchenko, V.} \& \bibinfo{author}{Wolynes, P.~G.}
\newblock \bibinfo{title}{Theory of aging in structural glasses}.
\newblock \emph{\bibinfo{journal}{J. Chem. Phys.}}
  \textbf{\bibinfo{volume}{121}}, \bibinfo{pages}{2852--2865}
  (\bibinfo{year}{2004}).

\bibitem{Ruta_JCP2013}
\bibinfo{author}{Ruta, B.}, \bibinfo{author}{Baldi, G.},
  \bibinfo{author}{Monaco, G.} \& \bibinfo{author}{Chushkin, Y.}
\newblock \bibinfo{title}{Compressed correlation functions and fast aging
  dynamics in metallic glasses}.
\newblock \emph{\bibinfo{journal}{J. Chem. Phys.}}
  \textbf{\bibinfo{volume}{138}}, \bibinfo{pages}{054508}
  (\bibinfo{year}{2013}).

\bibitem{ruta2013b}
\bibinfo{author}{Ruta, B.} \emph{et~al.}
\newblock \bibinfo{title}{Relaxation dynamics and aging in structural glasses}.
\newblock \emph{\bibinfo{journal}{AIP Conf. Proc.}}
  \textbf{\bibinfo{volume}{1518}}, \bibinfo{pages}{181} (\bibinfo{year}{2013}).

\bibitem{pineda2013}
\bibinfo{author}{Pineda, E.}, \bibinfo{author}{Bruna, P.},
  \bibinfo{author}{Ruta, B.}, \bibinfo{author}{Gonzalez-Silveira, M.} \&
  \bibinfo{author}{Crespo, D.}
\newblock \bibinfo{title}{Relaxation of rapidly quenched metallic glasses:
  Effect of the relaxation state on the slow low temperature dynamics}.
\newblock \emph{\bibinfo{journal}{Acta Mater.}} \textbf{\bibinfo{volume}{61}},
  \bibinfo{pages}{3002} (\bibinfo{year}{2013}).

\bibitem{pineda2010}
\bibinfo{author}{Pineda, E.}, \bibinfo{author}{Serrano, J.},
  \bibinfo{author}{Bruna, P.} \& \bibinfo{author}{Crespob, D.}
\newblock \bibinfo{title}{Fragility measurement of pd-based metallic glass by
  dynamic mechanical analysis}.
\newblock \emph{\bibinfo{journal}{J. Alloy and Comp.}}
  \textbf{\bibinfo{volume}{504}}, \bibinfo{pages}{S215} (\bibinfo{year}{2010}).

\bibitem{cavagna2009}
\bibinfo{author}{Cavagna, A.}
\newblock \bibinfo{title}{Supercooled liquids for pedestrians}.
\newblock \emph{\bibinfo{journal}{Phys. Rep.}} \textbf{\bibinfo{volume}{476}},
  \bibinfo{pages}{51} (\bibinfo{year}{2009}).

\bibitem{ruta2012b}
\bibinfo{author}{Ruta, B.} \emph{et~al.}
\newblock \bibinfo{title}{Acoustic exicitations in glassy sorbitol and their
  relation with the fragility and the boson peak}.
\newblock \emph{\bibinfo{journal}{J. Chem. Phys.}}
  \textbf{\bibinfo{volume}{137}}, \bibinfo{pages}{214502}
  (\bibinfo{year}{2012}).

\bibitem{Hachenberg_ApplPhysLett2008}
\bibinfo{author}{Hachenberg, J.} \emph{et~al.}
\newblock \bibinfo{title}{Merging of the $\alpha$ and $\beta$ relaxations and
  aging via the johari-goldstein modes in rapidly quenched metallic glasses}.
\newblock \emph{\bibinfo{journal}{Appl. Phys. Lett.}}
  \textbf{\bibinfo{volume}{92}}, \bibinfo{pages}{131911}
  (\bibinfo{year}{2008}).

\bibitem{orsi2014}
\bibinfo{author}{Orsi, D.} \emph{et~al.}
\newblock \bibinfo{title}{Controlling the dynamics of a bidimensional gel above
  and below its percolation transition}.
\newblock \emph{\bibinfo{journal}{Phys. Rev. E}} \textbf{\bibinfo{volume}{89}},
  \bibinfo{pages}{042308} (\bibinfo{year}{2014}).

\bibitem{guo2011}
\bibinfo{author}{Guo, H.}, \bibinfo{author}{Ramakrishnan, S.},
  \bibinfo{author}{Harden, J.~L.} \& \bibinfo{author}{Leheny, R.~L.}
\newblock \bibinfo{title}{Gel formation and aging in weakly attractive
  nanocolloid suspensions at intermediate concentrations}.
\newblock \emph{\bibinfo{journal}{J. Chem. Phys.}}
  \textbf{\bibinfo{volume}{135}}, \bibinfo{pages}{154903}
  (\bibinfo{year}{2011}).

\bibitem{czakkel2011}
\bibinfo{author}{Czakkel, O.} \& \bibinfo{author}{Madsen, A.}
\newblock \bibinfo{title}{Evolution of dynamics and structure during formation
  of a cross-linked polymer gel}.
\newblock \emph{\bibinfo{journal}{Europhys. Lett.}}
  \textbf{\bibinfo{volume}{95}}, \bibinfo{pages}{28001} (\bibinfo{year}{2011}).

\bibitem{sidebottom1993}
\bibinfo{author}{Sidebottom, D.}, \bibinfo{author}{Bergman, R.},
  \bibinfo{author}{Borjesson, L.} \& \bibinfo{author}{Torell, L.~M.}
\newblock \bibinfo{title}{Two-step relaxation decay in a strong glass former}.
\newblock \emph{\bibinfo{journal}{Phys. Rev. Lett.}}
  \textbf{\bibinfo{volume}{71}}, \bibinfo{pages}{2260} (\bibinfo{year}{1993}).

\bibitem{mezei1987}
\bibinfo{author}{Mezei, F.}, \bibinfo{author}{Knaak, W.} \&
  \bibinfo{author}{Farago, B.}
\newblock \bibinfo{title}{Neutron spin-echo study of dynamic correlations near
  the liquid-glass transition}.
\newblock \emph{\bibinfo{journal}{Phys. Rev. Lett.}}
  \textbf{\bibinfo{volume}{58}}, \bibinfo{pages}{571} (\bibinfo{year}{1987}).

\bibitem{ngaibook}
\bibinfo{author}{Ngai, K.}
\newblock \emph{\bibinfo{title}{Relaxation and Diffusion in Complex Systems}}
  (\bibinfo{publisher}{Springer}, \bibinfo{year}{2011}).

\bibitem{johari1970}
\bibinfo{author}{Johari, G.~P.} \& \bibinfo{author}{Goldstein, M.}
\newblock \bibinfo{title}{Viscous liquids and the glass transition. ii.
  secondary relaxations in glasses of rigid molecules}.
\newblock \emph{\bibinfo{journal}{J. Chem. Phys.}}
  \textbf{\bibinfo{volume}{53}}, \bibinfo{pages}{2372} (\bibinfo{year}{1970}).

\bibitem{Yu_MatTod2013}
\bibinfo{author}{Yu, H.~B.}, \bibinfo{author}{Wang, W.~H.} \&
  \bibinfo{author}{Samwer, K.}
\newblock \bibinfo{title}{The $\beta$ relaxation in metallic glasses: an
  overview}.
\newblock \emph{\bibinfo{journal}{Materials Today}}
  \textbf{\bibinfo{volume}{16}}, \bibinfo{pages}{183--190}
  (\bibinfo{year}{2013}).

\bibitem{Liu_NatComm2014}
\bibinfo{author}{Liu, Y.}, \bibinfo{author}{Fujita, T.}, \bibinfo{author}{Aji,
  D.}, \bibinfo{author}{Matsuura, M.} \& \bibinfo{author}{Chen, M.}
\newblock \bibinfo{title}{Structural origins of johari-goldstein relaxation in
  a metallic glass}.
\newblock \emph{\bibinfo{journal}{Nat. Commun.}} \textbf{\bibinfo{volume}{5}},
  \bibinfo{pages}{3238} (\bibinfo{year}{2014}).

\bibitem{wang2014}
\bibinfo{author}{Wang, Q.} \emph{et~al.}
\newblock \bibinfo{title}{Unusual fast secondary relaxation in metallic glass}.
\newblock \emph{\bibinfo{journal}{Nat. Comm.}} \textbf{\bibinfo{volume}{6}},
  \bibinfo{pages}{7876} (\bibinfo{year}{2015}).

\bibitem{Ruta_JAllComp2014}
\bibinfo{author}{Ruta, B.}, \bibinfo{author}{Giordano, V.~M.},
  \bibinfo{author}{Erra, L.}, \bibinfo{author}{Liu, C.} \&
  \bibinfo{author}{Pineda, E.}
\newblock \bibinfo{title}{Structural and dynamical properties of Mg65Cu25Y10
  metallic glasses studied by in situ high energy x-ray diffraction and time
  resolved x-ray photon correlation spectroscopy}.
\newblock \emph{\bibinfo{journal}{J. Alloys Compd.}}
  \textbf{\bibinfo{volume}{615}}, \bibinfo{pages}{S45--S50}
  (\bibinfo{year}{2014}).

\bibitem{Mattern_ActaMat2012}
\bibinfo{author}{Mattern, N.}, \bibinfo{author}{Stoica, M.},
  \bibinfo{author}{Vaughan, G.} \& \bibinfo{author}{Eckert, J.}
\newblock \bibinfo{title}{Thermal behaviour of
  pd$_{40}$cu$_{30}$ni$_{10}$p$_{20}$ bulk metallic glass}.
\newblock \emph{\bibinfo{journal}{Acta Mat.}} \textbf{\bibinfo{volume}{60}},
  \bibinfo{pages}{517--524} (\bibinfo{year}{2012}).

\bibitem{Qian_MatSciEng1991}
\bibinfo{author}{Fu-Qian, Z.}
\newblock \bibinfo{title}{Thermal expansion and fractional free volume changes
  of metallic glasses during heating}.
\newblock \emph{\bibinfo{journal}{Materials Science and Engineering A}}
  \textbf{\bibinfo{volume}{A134}}, \bibinfo{pages}{998--999}
  (\bibinfo{year}{1991}).

\bibitem{Ma_NatMat2009}
\bibinfo{author}{Ma, D.}, \bibinfo{author}{Stoica, A.~D.} \&
  \bibinfo{author}{Wang, X.-L.}
\newblock \bibinfo{title}{Power-law scaling and fractal nature of medium-range
  order in metallic glasses}.
\newblock \emph{\bibinfo{journal}{Nature Mater.}} \textbf{\bibinfo{volume}{8}},
  \bibinfo{pages}{30--34} (\bibinfo{year}{2009}).

\bibitem{Sokolov_PRL1992}
\bibinfo{author}{Sokolov, A.~P.}, \bibinfo{author}{Kisliuk, A.},
  \bibinfo{author}{Soltwisch, M.} \& \bibinfo{author}{Quitmann, D.}
\newblock \bibinfo{title}{Medium-range order in glasses: Comparison of raman
  and diffraction measurements}.
\newblock \emph{\bibinfo{journal}{Phys. Rev. Lett.}}
  \textbf{\bibinfo{volume}{69}}, \bibinfo{pages}{1540} (\bibinfo{year}{1992}).

\bibitem{greaves2007}
\bibinfo{author}{Greaves, G.} \& \bibinfo{author}{Sen, S.}
\newblock \bibinfo{title}{Inorganic glasses, glass-forming liquids and
  amorphizing solids}.
\newblock \emph{\bibinfo{journal}{Adv. Phys.}} \textbf{\bibinfo{volume}{56}},
  \bibinfo{pages}{1} (\bibinfo{year}{2007}).

\bibitem{debenedetti2001}
\bibinfo{author}{Debenedetti, P.~G.} \& \bibinfo{author}{Stillinger, F.~H.}
\newblock \bibinfo{title}{Supercooled liquids and the glass transition}.
\newblock \emph{\bibinfo{journal}{Nature}} \textbf{\bibinfo{volume}{410}},
  \bibinfo{pages}{259} (\bibinfo{year}{2001}).

\bibitem{rodney2011}
\bibinfo{author}{Rodney, D.}, \bibinfo{author}{Tanguy, A.} \&
  \bibinfo{author}{Vandembroucq, D.}
\newblock \bibinfo{title}{Modeling the mechanics of amorphous solids at
  different length scale and time scale}.
\newblock \emph{\bibinfo{journal}{Modelling Simul. Mater. Sci. Eng.}}
  \textbf{\bibinfo{volume}{19}}, \bibinfo{pages}{083001}
  (\bibinfo{year}{2011}).

\bibitem{heuer2008}
\bibinfo{author}{Heuer, A.}
\newblock \bibinfo{title}{Exploring the potential energy landscape of
  glass-forming systems: from inherent structures via metabasins to macroscopic
  transport}.
\newblock \emph{\bibinfo{journal}{J. Phys. Cond. Matter}}
  \textbf{\bibinfo{volume}{20}}, \bibinfo{pages}{373101}
  (\bibinfo{year}{2008}).

\bibitem{fan2014}
\bibinfo{author}{Fan, Y.}, \bibinfo{author}{Iwashita, T.} \&
  \bibinfo{author}{Egami, T.}
\newblock \bibinfo{title}{How thermally activated deformation starts in
  metallic glass}.
\newblock \emph{\bibinfo{journal}{Nat. Comm.}} \textbf{\bibinfo{volume}{5}},
  \bibinfo{pages}{5083} (\bibinfo{year}{2014}).

\bibitem{fan2015}
\bibinfo{author}{Fan, Y.}, \bibinfo{author}{Iwashita, T.} \&
  \bibinfo{author}{Egami, T.}
\newblock \bibinfo{title}{Crossover from localized to cascade relaxations in
  metallic glasses}.
\newblock \emph{\bibinfo{journal}{Phys. Rev. Lett.}}
  \textbf{\bibinfo{volume}{115}}, \bibinfo{pages}{045501}
  (\bibinfo{year}{2015}).

\bibitem{samwer2014}
\bibinfo{author}{Krisponeit, J.~O.} \emph{et~al.}
\newblock \bibinfo{title}{Crossover from random three-dimensional avalanches to
  correlated nano shear bands in metallic glasses}.
\newblock \emph{\bibinfo{journal}{Nat. Comm.}} \textbf{\bibinfo{volume}{5}},
  \bibinfo{pages}{3616} (\bibinfo{year}{2014}).

\bibitem{wang2014Natcom}
\bibinfo{author}{Wang, Z.}, \bibinfo{author}{Sun, B.~A.}, \bibinfo{author}{Bai,
  H.~Y.} \& \bibinfo{author}{Wang, W.~H.}
\newblock \bibinfo{title}{Evolution of hidden localized flow during
  glass-to-liquid transition in metallic glass}.
\newblock \emph{\bibinfo{journal}{Nat. Comm.}} \textbf{\bibinfo{volume}{5}},
  \bibinfo{pages}{5823} (\bibinfo{year}{2014}).

\bibitem{busch1998b}
\bibinfo{author}{Busch, R.}, \bibinfo{author}{Liu, W.} \&
  \bibinfo{author}{Johnson, W.~L.}
\newblock \bibinfo{title}{Thermodynamics and kinetics of the
  mg$_{65}$cu$_{25}$y$_{10}$ bulk metallic glass forming liquid}.
\newblock \emph{\bibinfo{journal}{J. Appl. Phys.}}
  \textbf{\bibinfo{volume}{83}}, \bibinfo{pages}{4134--4141}
  (\bibinfo{year}{1998}).

\bibitem{Kumar2013}
\bibinfo{author}{Kumar, G.}, \bibinfo{author}{Neibecker, P.},
  \bibinfo{author}{Liu, Y.~H.} \& \bibinfo{author}{Schroers, J.}
\newblock \bibinfo{title}{Critical fictive temperature for plasticity in
  metallic glasses}.
\newblock \emph{\bibinfo{journal}{Nat. Comm.}} \textbf{\bibinfo{volume}{4}},
  \bibinfo{pages}{1536} (\bibinfo{year}{2013}).

\bibitem{Kumar2009}
\bibinfo{author}{Kumar, G.}, \bibinfo{author}{Rector, D.},
  \bibinfo{author}{Conner, R.} \& \bibinfo{author}{Schroers, J.}
\newblock \bibinfo{title}{Embrittlement of zr-based bulk metallic glasses}.
\newblock \emph{\bibinfo{journal}{Acta Mat.}} \textbf{\bibinfo{volume}{57}},
  \bibinfo{pages}{3572} (\bibinfo{year}{2009}).

\bibitem{Kumar2011b}
\bibinfo{author}{Kumar, G.}, \bibinfo{author}{Prades-Rodel, S.},
  \bibinfo{author}{Blatter, A.} \& \bibinfo{author}{Schroers, J.}
\newblock \bibinfo{title}{Unusual brittle behavior of pd-based bulk metallic
  glass}.
\newblock \emph{\bibinfo{journal}{Scripta Materialia}}
  \textbf{\bibinfo{volume}{65}}, \bibinfo{pages}{585} (\bibinfo{year}{2011}).

\bibitem{Liu_Nature1998}
\bibinfo{author}{Liu, A.~J.} \& \bibinfo{author}{Nagel, S.~R.}
\newblock \bibinfo{title}{Jamming is not just cool any more}.
\newblock \emph{\bibinfo{journal}{Nature}} \textbf{\bibinfo{volume}{396}},
  \bibinfo{pages}{21} (\bibinfo{year}{1998}).

\bibitem{Chushkin2012}
\bibinfo{author}{Chushkin, Y.}, \bibinfo{author}{Caronna, C.} 
\& \bibinfo{author}{Madsen, A.}
\newblock \bibinfo{title}{A novel event correlation scheme for
X-ray photon correlation spectroscopy}.
\newblock \emph{\bibinfo{journal}{J. Appl. Crystallogr.}} \textbf{\bibinfo{volume}{45}},
  \bibinfo{pages}{807} (\bibinfo{year}{2012}).

\bibitem{Eggert2002}
\bibinfo{author}{Eggert, J.~H.}, \bibinfo{author}{Weck, G.},
  \bibinfo{author}{Loubeyre, P.} \& \bibinfo{author}{Mezouar, M.}
\newblock \bibinfo{title}{Quantitative structure factor and density
  measurements of high-pressure fluids in diamond anvil cells by x-ray
  diffraction: Argon and water}.
\newblock \emph{\bibinfo{journal}{Phys. Rev. B}} \textbf{\bibinfo{volume}{65}},
  \bibinfo{pages}{174105} (\bibinfo{year}{2002}).

\end{thebibliography}

\newpage
\textbf{Aknowledgements}
The authors thank K. L'Hoste and H. Vitoux for technical support, E. Pineda and P. Bruna for the sample preparation and S. J. A. Kimber, M. Di Michiel and G. Baldi for help during the experiments. Y. Chushkin is gratefully thank for the assistance during the XPCS measurements and for providing the software for the analysis of the XPCS data.
\\

\textbf{Contributions}
V.M.G. and B.R. conceived the project and performed all experiments. B.R. analyzed the XPCS data, V.M.G. analyzed the diffraction data. Both authors discussed the data and wrote together the manuscript. 
\\

\textbf{Competing financial interests}
The authors declare no competing financial interests
\\

\textbf{Figure Legends} \\

\textbf{Figure 1} \\
\textbf{Temperature and Age dependence of the relaxation dynamics} (a): Temporal evolution of intensity-intensity correlation functions measured with XCPS for $q_0=2.81$ \AA$^{-1}$ during isotherm at $T=453 K$. Data are reported after baseline subtraction together with the fits obtained by the KWW model (see text). The arrow indicates the evolution with annealing time, $t_a$ from temperature equilibration. (b): Temporal and $T$ evolution of the structural relaxation time measured with XPCS as a function of $t-t_0$, where $t_0$ is the time corresponding to the beginning of the heating protocol. (c): Corresponding shape parameters. (d): Selection of intensity correlation functions measured at fixed $t_a$ and different $T$ reported as a function of $t/\tau_{\alpha}(t_a,T)$. (e): Temporal and $T$ evolution of the initial plateau value,  $c f_q(T)^2$, as a function of $t-t_0$. In (b),(c),(d) and (e)  $T$ steps are: 393~K (blue empty circles), 433~K (green full circles), 453~K (red triangles), 473~K (cyan down triangles), 493~K (purple squares), 513~K (black diamonds). \\

\textbf{Figure 2} \\
{\textbf{Decay of the intensity correlation function in presence of a secondary relaxation}. We report here the typical behavior of the intensity correlation function as measured by XPCS in presence of a secondary relaxation process. Instead that a single decay (black dashed line), the $g_2$ displays a first step at very short time scales associated to the secondary $\beta$-process and a second decay at much larger time scales due to the main structural relaxation process. The colored region in the Figure represents the dynamical range investigated by XPCS, which starts from $t>3$s. Here, the activation of a faster relaxation cannot be directly observed but it leads to a decrease of the initial plateau of an amount that corresponds to the strength, $\Delta[c*f_q^2]$, of the fast $\beta$-process.\\

\textbf{Figure 3} \\
\textbf{Structural characterization during aging and thermal treatment} Static structure factor S(q) after background subtraction and normalization (panel a), and correlation function G(r) (panel b) are reported as a function of temperature, at the beginning of each isotherm, right after temperature equilibration. Curves labels are for different temperatures: a=298~K, b=393~K,c=433~K,d=453~K,e=473~K,f=493~K,g=513~K.    \\

\textbf{Figure 4} \\
\textbf{Aging, volume reduction and medium range ordering} (a): Relative volume change (see text), reported as a function of $t-t_0$. Each temperature ramp  (crosses) is followed by an isotherm. Symbols are as in Fig~\ref{Fig1}. Inset: zoom of data at 433~K reported as a function of time from the beginning of the isotherm, together with the best fit to an exponential law. (b): Relative volume change as a function of temperature. Lines: temperature ramps; empty circles, empty squares: first and  last point of an isotherm respectively. For a better visibility, we plot only data for $T \geq 373$~K. (c): Temporal evolution of the relative change of the width of the FSDP. (d): Characteristic time for aging as obtained from the volume relaxation ($\tau_{\rm V}$,blue empty squares), the narrowing of the FSDP ($\tau_{{\rm \Gamma}}$, blue empty circles) and from XPCS data ($\tau*$, black full dots). The average of $\tau_{\rm V}$ and $\tau_{\rm {\Gamma}}$ is also reported (red empty circles). Error bars correspond to the error of the fitting procedure and to error propagation for the average of $\tau_{\rm V}$ and $\tau_{\rm {\Gamma}}$. Where not shown, error bars are within the symbol size.\\

\textbf{Figure 5} \\
\textbf{Atomic rearrangements behind the secondary fast relaxation process} Relative change of position (panels a,b,c) and width (panels d,e,f) of the first three shells of G(r) as a function of time during the different isotherms. $r_{\rm 1}$, r$_{\rm 2}$ and $r_{\rm 4}$ correspond to the positions reported in Fig.~\ref{Fig2}.  Symbols are as in Fig~\ref{Fig1}. Error bars for positions are within the symbol size.\\

\textbf{Figure 6} \\
\textbf{Sketch of the results} (a): Behavior of a correlation function measured with XPCS; Evolution of the structure and the macroscopic properties during the dynamical (b) and stationary (c) aging regimes measured with XPCS. Explanation in the text.\\
\\
\newpage
\begin{figure}
  \centering \includegraphics[width=\columnwidth]{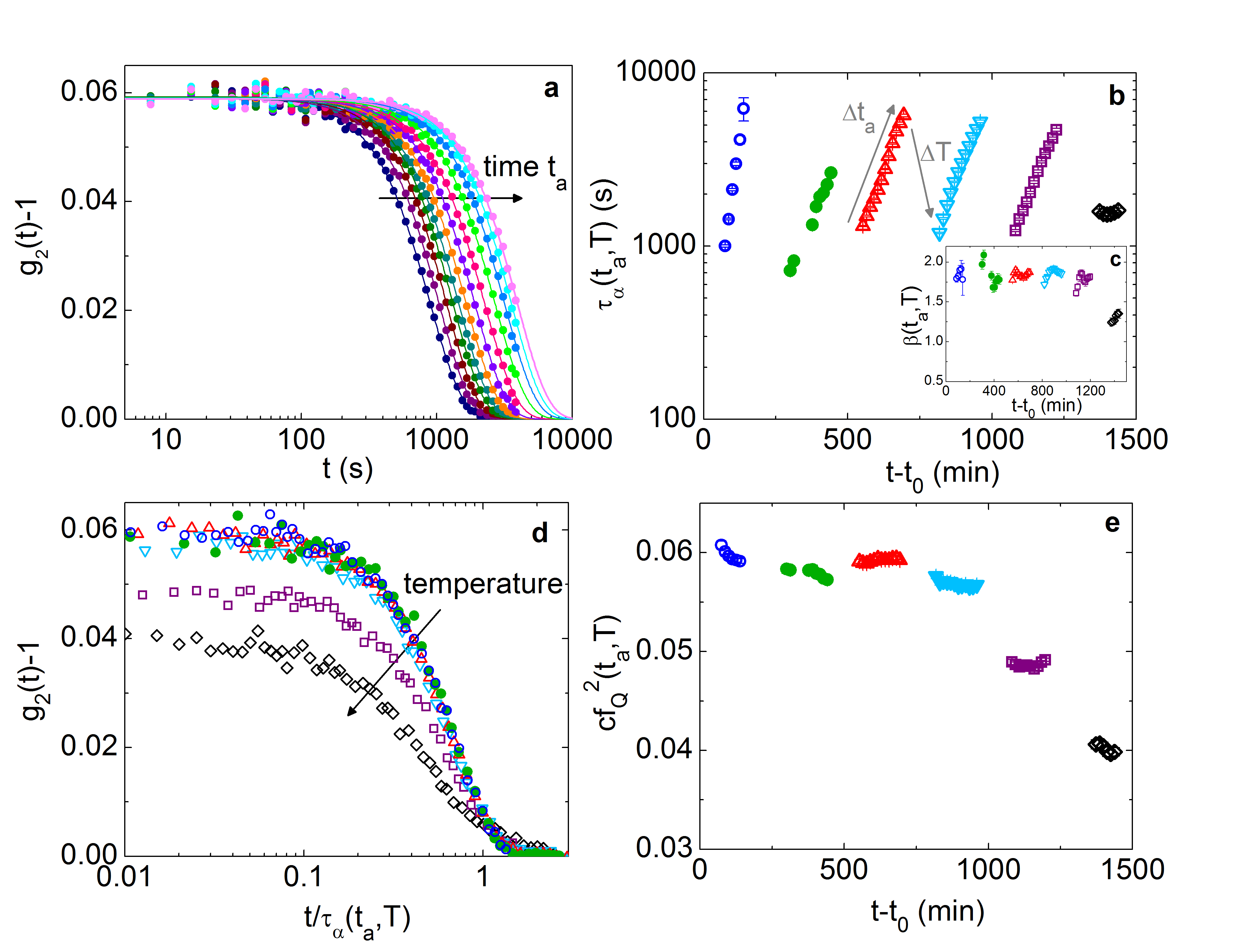}
	\vspace{-0.8cm}
  \caption{\textbf{Temperature and Age dependence of the relaxation dynamics} (a): Temporal evolution of intensity-intensity correlation functions measured with XCPS for $q_0=2.81$ \AA$^{-1}$ during isotherm at $T=453 K$. Data are reported after baseline subtraction together with the fits obtained by the KWW model (see text). The arrow indicates the evolution with annealing time, $t_a$ from temperature equilibration. (b): Temporal and $T$ evolution of the structural relaxation time measured with XPCS as a function of $t-t_0$, where $t_0$ is the time corresponding to the beginning of the heating protocol. (c): Corresponding shape parameters. (d): Selection of intensity correlation functions measured at fixed $t_a$ and different $T$ reported as a function of $t/\tau_{\alpha}(t_a,T)$. (e): Temporal and $T$ evolution of the initial plateau value,  $c f_q(T)^2$, as a function of $t-t_0$. In (b),(c),(d) and (e)  $T$ steps are: 393~K (blue empty circles), 433~K (green full circles), 453~K (red triangles), 473~K (cyan down triangles), 493~K (purple squares), 513~K (black diamonds). }
\label{Fig1}
\vspace{-0.5cm}
\end{figure}

\begin{figure}
  \centering \includegraphics[width=\columnwidth]{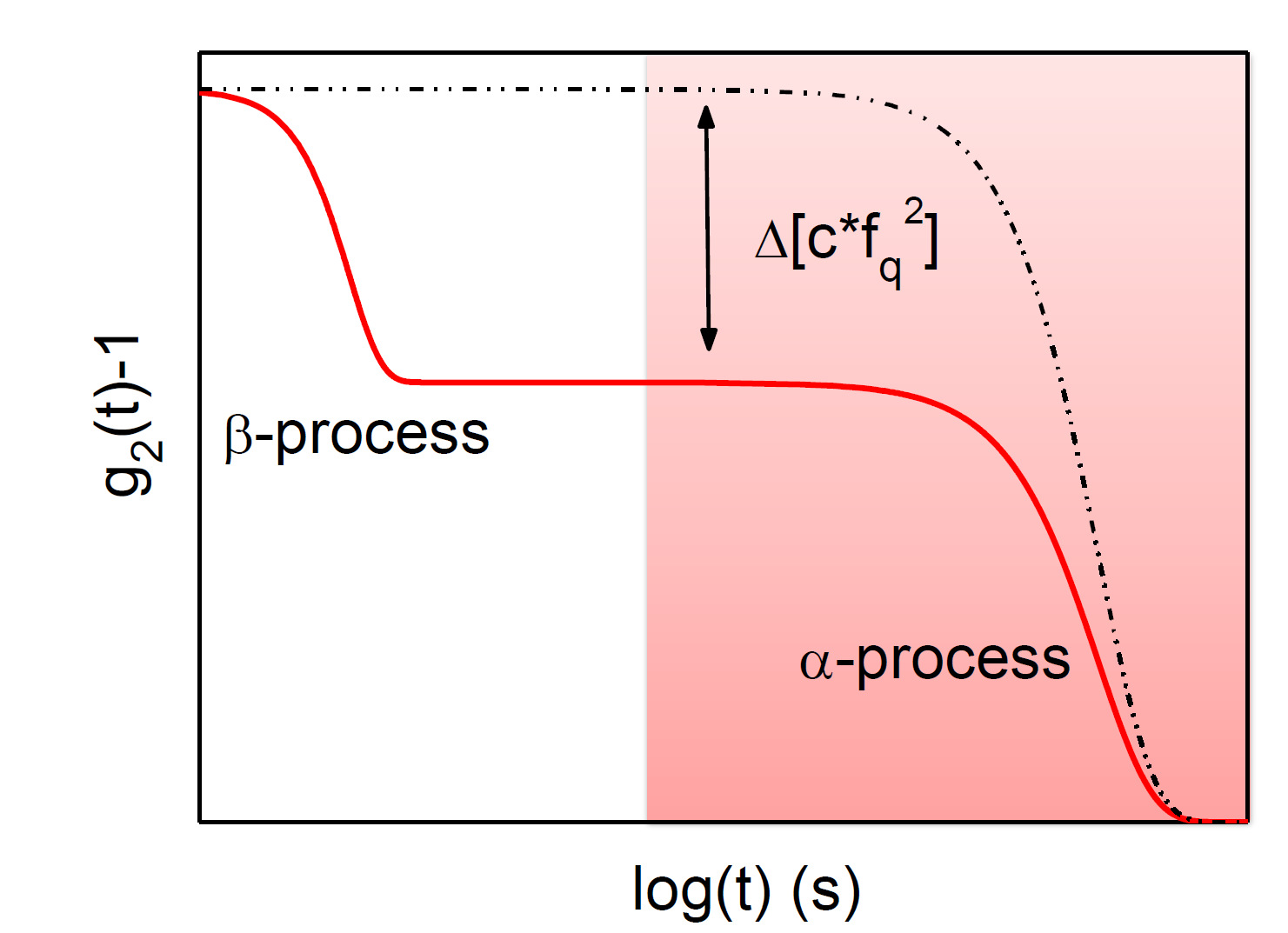}
	\vspace{-0.8cm}
  \caption{\textbf{Decay of the intensity correlation function in presence of a secondary relaxation}. We report here the typical behavior of the intensity correlation function as measured by XPCS in presence of a secondary relaxation process.  Instead that a single decay (black dashed line), the $g_2$ displays a first step at very short time scales associated to the secondary $\beta$-process and a second decay at much larger time scales due to the main structural relaxation process. The colored region in the Figure represents the dynamical range investigated by XPCS, which starts from $t>3$s. Here, the activation of a faster relaxation cannot be directly observed but it leads to a decrease of the initial plateau of an amount that corresponds to the strength, $\Delta[c*f_q^2]$, of the fast $\beta$-process.}
\label{Fig2XPCS}
\end{figure}

\begin{figure}
	\centering \includegraphics[width=\columnwidth]{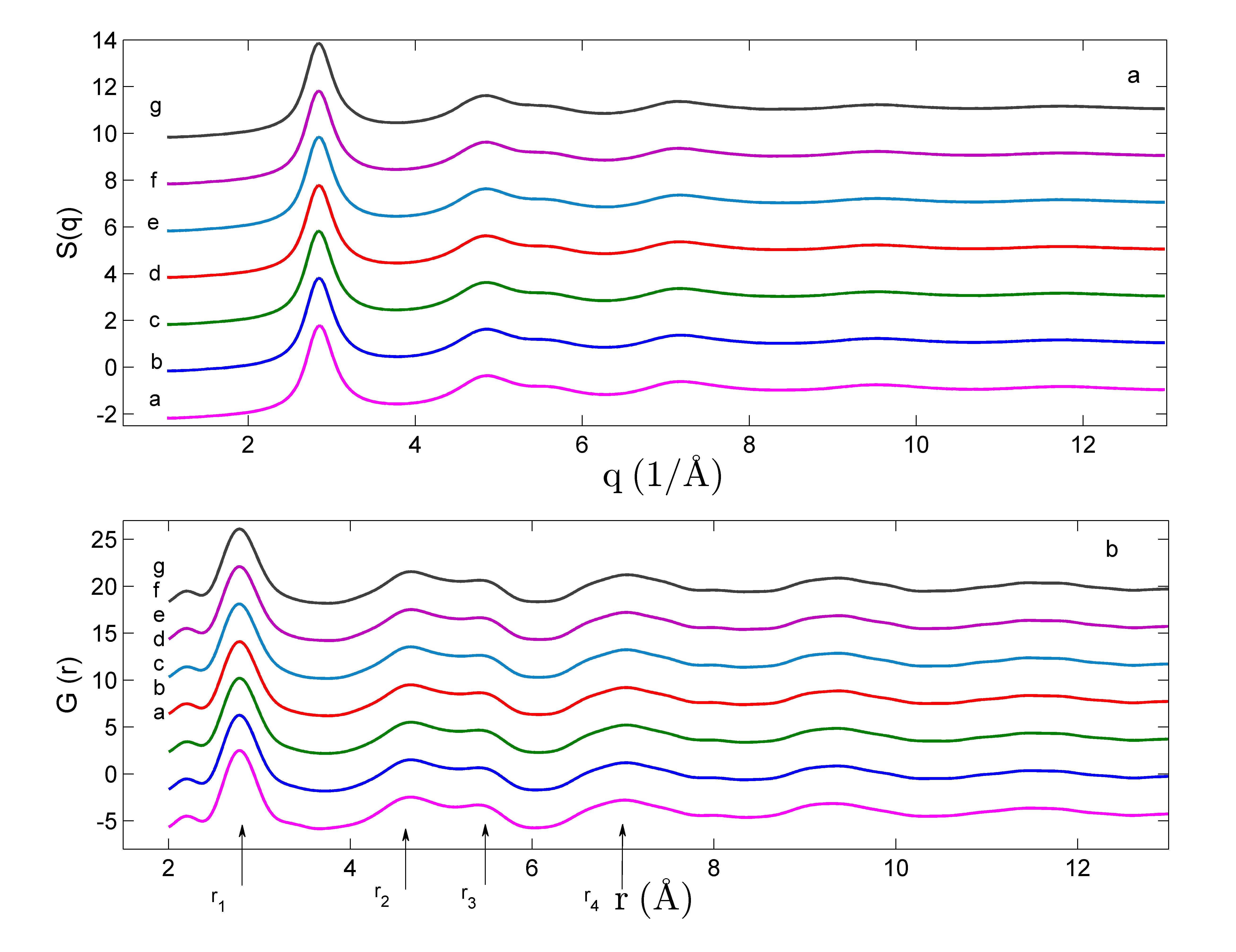}
	\vspace{-0.3cm}
  \caption{\textbf{Structural characterization during aging and thermal treatment} Static structure factor S(q) after background subtraction and normalization (panel a), and correlation function G(r) (panel b) are reported as a function of temperature, at the beginning of each isotherm, right after temperature equilibration. Curves labels are for different temperatures: a=298~K, b=393~K,c=433~K,d=453~K,e=473~K,f=493~K,g=513~K.  }
\label{Fig2}
\vspace{-0.5cm}
\end{figure}

\begin{figure}
  \centering \includegraphics[width=\columnwidth]{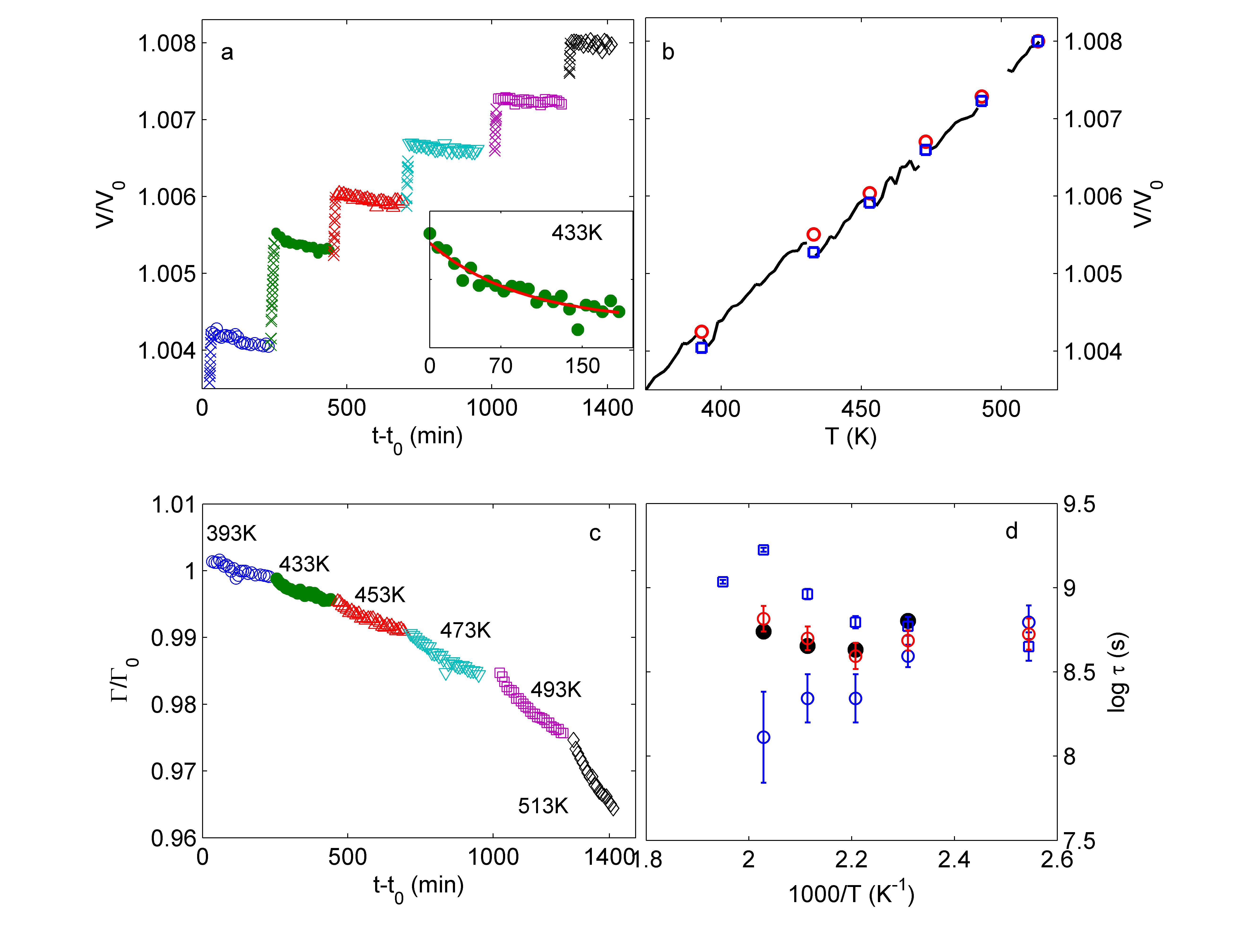}
	\vspace{-0.8cm}
  \caption{\textbf{Aging, volume reduction and medium range ordering} (a): Relative volume change (see text), reported as a function of $t-t_0$. Each temperature ramp  (crosses) is followed by an isotherm. Symbols are as in Fig~\ref{Fig1}. Inset: zoom of data at 433~K reported as a function of time from the beginning of the isotherm, together with the best fit to an exponential law. (b): Relative volume change as a function of temperature. Lines: temperature ramps; empty circles, empty squares: first and  last point of an isotherm respectively. For a better visibility, we plot only data for $T \geq 373$~K. (c): Temporal evolution of the relative change of the width of the FSDP. (d): Characteristic time for aging as obtained from the volume relaxation ($\tau_{\rm V}$,blue empty squares), the narrowing of the FSDP ($\tau_{{\rm \Gamma}}$, blue empty circles) and from XPCS data ($\tau*$, black full dots). The average of $\tau_{\rm V}$ and $\tau_{\rm {\Gamma}}$ is also reported (red empty circles). Error bars correspond to the error of the fitting procedure and to error propagation for the average of $\tau_{\rm V}$ and $\tau_{\rm {\Gamma}}$. Where not shown, error bars are within the symbol size.}
\label{Fig3}
\vspace{-0.5cm}
\end{figure} 

\begin{figure}
  \centering \includegraphics[width=\columnwidth]{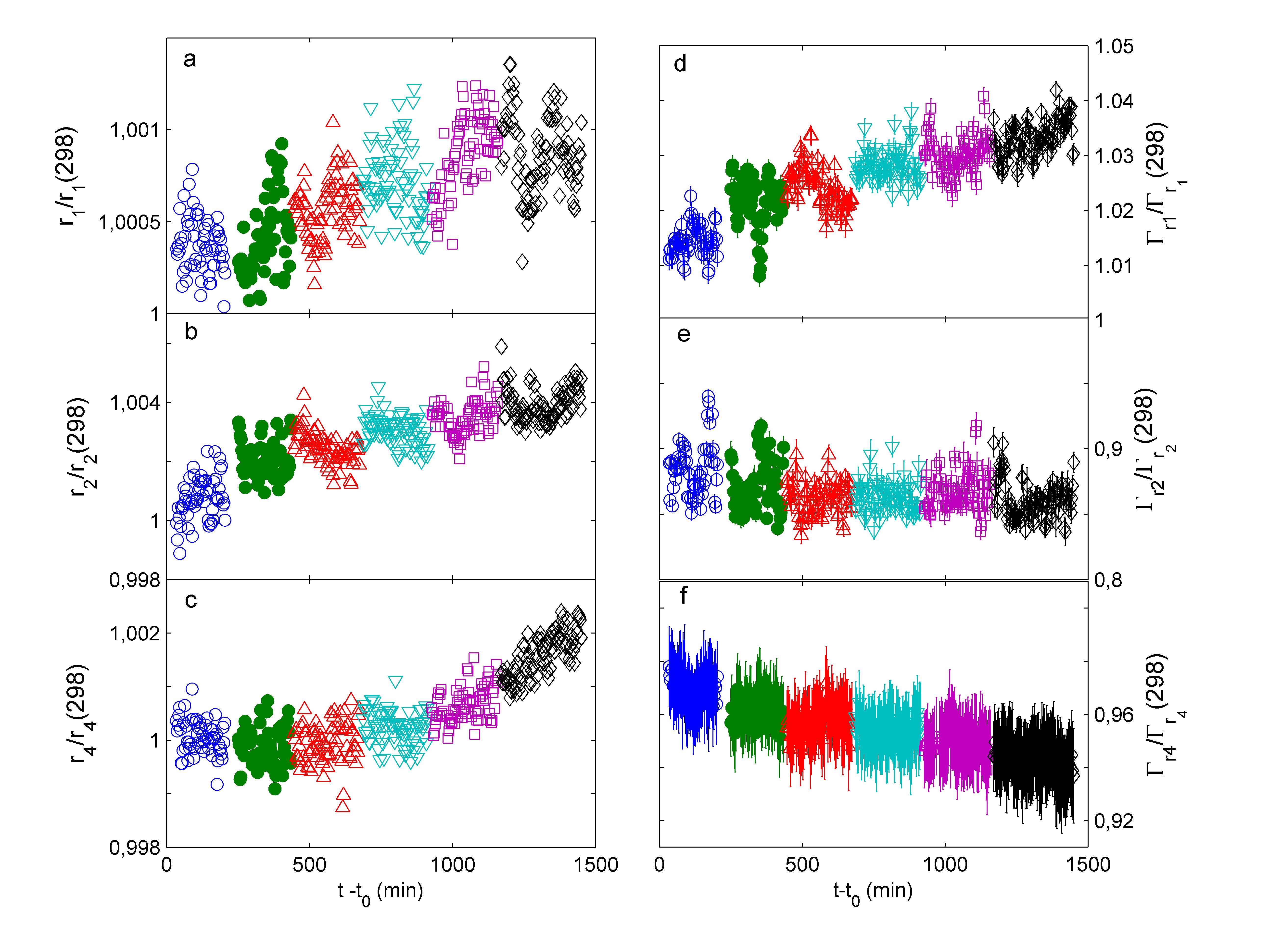}
	\vspace{-0.8cm}
  \caption{ \textbf{Atomic rearrangements behind the secondary fast relaxation process} Relative change of position (panels a,b,c) and width (panels d,e,f) of the first three shells of G(r) as a function of time during the different isotherms. $r_{\rm 1}$, r$_{\rm 2}$ and $r_{\rm 4}$ correspond to the positions reported in Fig.~\ref{Fig2}.  Symbols are as in Fig~\ref{Fig1}. Error bars for positions are within the symbol size.
	 }
\label{Fig4}
\vspace{-0.5cm}
\end{figure}

\begin{figure}
  \centering \includegraphics[width=\columnwidth]{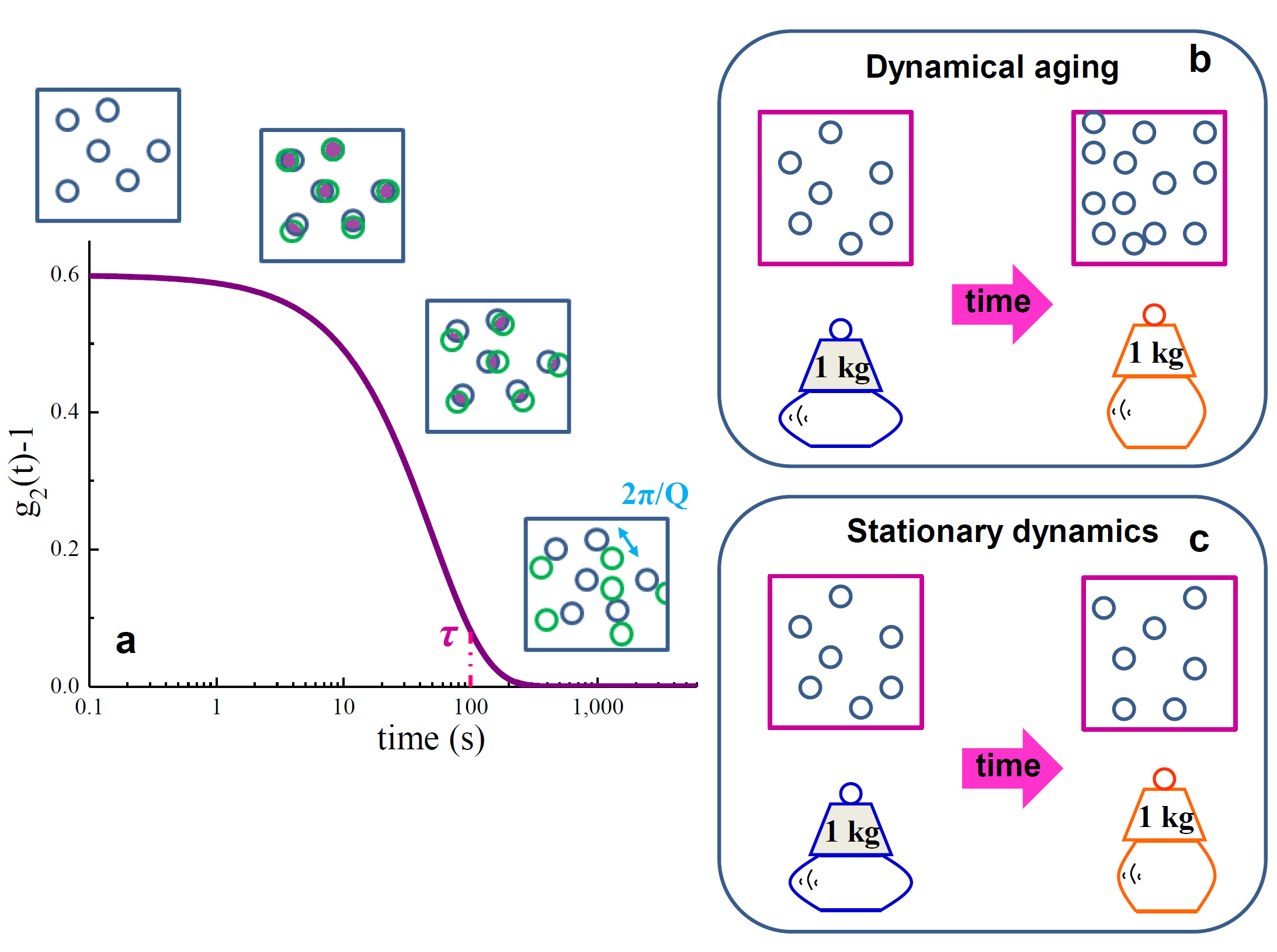}
	\vspace{-0.8cm}
  \caption{ \textbf{Sketch of the results} (a): Behavior of a correlation function measured with XPCS; Evolution of the structure and the macroscopic properties during the dynamical (b) and stationary (c) aging regimes measured with XPCS. Explanation in the text.
	 }
\label{Fig6}
\vspace{-0.5cm}
\end{figure}

\end{document}